\documentclass[12pt,preprint]{aastex}
\begin{document}

\baselineskip=0.8cm

\parindent=1.0cm

\title{The Star-Forming Histories of the Nucleus, Bulge, and Inner Disk of NGC 5102: 
Clues to the Evolution of a Nearby Lenticular Galaxy. 
\altaffilmark{1}\altaffilmark{2}\altaffilmark{3}}

\author{T. J. Davidge}

\affil{Dominion Astrophysical Observatory,
\\National Research Council of Canada, 5071 West Saanich Road,
\\Victoria, BC Canada V9E 2E7\\tim.davidge@nrc.ca}

\altaffiltext{1}{Based on observations obtained at the Gemini Observatory, which is
operated by the Association of Universities for Research in Astronomy, Inc., under a
cooperative agreement with the NSF on behalf of the Gemini partnership: the National
Science Foundation (United States), the National Research Council (Canada), CONICYT
(Chile), the Australian Research Council (Australia), Minist\'{e}rio da Ci\^{e}ncia,
Tecnologia e Inova\c{c}\~{a}o (Brazil) and Ministerio de Ciencia, Tecnolog\'{i}a e
Innovaci\'{o}n Productiva (Argentina).}

\altaffiltext{2}{This research used the facilities of the Canadian Astronomy Data
Centre operated by the National Research Council of Canada with the support
of the Canadian Space Agency.}

\altaffiltext{3}{This research has made use of the NASA/IPAC Infrared Science Archive,
which is operated by the Jet Propulsion Laboratory, California Institute of Technology,
under contract with the National Aeronautics and Space Administration.}

\begin{abstract}

	Long slit spectra recorded with GMOS on Gemini South are used to examine 
the star-forming history of the lenticular galaxy NGC 5102. Structural and 
supplemental photometric information are obtained from archival Spitzer [3.6] images. 
Absorption features at blue and visible wavelengths are traced out 
along the minor axis to galactoentric radii $\sim 60$ arcsec ($\sim 0.9$ kpc), 
sampling the nucleus, bulge, and disk components. 
Comparisons with model spectra point to luminosity-weighted metallicities 
that are consistent with the colors of resolved red giant 
branch stars in the disk. The nucleus has a luminosity-weighted 
age at visible wavelengths of $\sim 1^{+0.2}_{-0.1}$ Gyr, and the integrated light is 
dominated by stars that formed over a time period of only 
a few hundred Myr. For comparison, the luminosity-weighted ages of the bulge and 
disk are $\sim 2^{+0.5}_{-0.2}$ Gyr and 10$^{+2}_{-2}$ Gyr, 
respectively. The $g'-[3.6]$ colors of the nucleus and bulge are 
consistent with the spectroscopically-based ages. In contrast to the 
nucleus, models that assume star-forming activity spanning 
many Gyr provide a better match to the spectra of the bulge and disk than 
simple stellar population models. Isophotes in the bulge have a disky 
shape, hinting that the bulge was assembled from material with significant 
rotational support. The star-forming histories of the bulge and disk are consistent 
with the bulge forming from the collapse of a long-lived bar, rather than from the 
collapse of a transient structure that formed as the result of a tidal interaction. 
It is thus suggested that the progenitor of NGC 5102 was a barred disk galaxy that 
morphed into a lenticular galaxy through the buckling of its bar.

\end{abstract}


\section{INTRODUCTION}

	NGC 5102 is a lenticular galaxy that is part of the Cen A group 
(Karachentsev et al. 2002). As one of the nearest lenticular galaxies, it is an 
important laboratory for investigating the origins of this galaxy type in 
group environments. As is typical of lenticular galaxies, there is only modest 
star-forming activity in the disk (Davidge 2010; Karachentsev et al. 2002), and the 
present-day star formation rate (SFR) is $\sim 0.02$ M$_{\odot}$ year$^{-1}$ 
(Davidge 2008). The current star-forming activity is largely 
confined to the south east quadrant of the galaxy, where 5 of the 7 known 
HII regions are located (McMillan et al. 1994). 

	With a total stellar mass of $\sim 7 \times 10^9$ M$_{\odot}$ (Davidge 
2008), the SFR averaged over a Hubble time is $\sim 0.7$ M$_{\odot}$ year$^{-1}$, 
and so it is not surprising that that there are signs that the disk of NGC 5102 
experienced past levels of star-forming activity at rates that far surpass those seen 
today. Davidge (2008) investigated the properties of AGB stars in 
NGC 5102 and concluded that $\sim 20\%$ of the disk mass formed within the past 
Gyr, which is twice the pace that would be expected if the galaxy had experienced a 
constant SFR throughout its entire lifetime. The number density of C stars provides 
additional evidence of an elevated SFR during intermediate epochs (Davidge 2010). 
The decline in the disk SFR during recent epochs has been charted by 
Beaulieu et al. (2010), who investigated the star-forming history (SFH) of the inner 
disk of NGC 5102 (their F3). They find that the mean SFR between 40 
and 120 Myr in the past was only one third that 120 - 200 Myr in the 
past, while the SFR during the past 40 Myr has been negligible. 
NGC 5102 has a high HI gas mass when compared with other lenticular galaxies 
(van Woerden et al. 1993), and it is possible that this reservoir of cool gas may 
eventually re-kindle global star-forming activity.

	The stellar content of the nucleus of NGC 5102 also indicates that this 
was an area of high levels of star-forming activity during intermediate epochs. 
The nucleus has a blue color at visible wavelengths (Pritchet 
1979), and Bica (1988) finds that the central SFR peaked 0.5 Gyr in the 
past based on the integrated visible/red spectrum. Deharveng et al. (1997) 
conclude that large scale star-forming activity near the center of NGC 5102 
commenced 0.5 - 0.8 Gyr in the past, while Miner et al. (2011) find a 
luminosity-weighted age $\sim 0.3$ Gyr from near-infrared line indices. 

	Some structural properties of NGC 5102 hint at 
a past interaction with another system. The disk is warped, and there is a plume 
of stars that extends out to 16 disk scale lengths (Davidge 2010) that might be a 
tidal feature. While NGC 5102 does not have companions at the present day, a flyby 
encounter with a perturber that is now long gone could have left 
long-lasting signatures in the disk (Kim et al. 2014). 
Alternatively, if NGC 5102 was disturbed by a dwarf companion 
then that system may have been disrupted or consumed by NGC 5102. 
Beaulieu et al. (2010) suggest that NGC 5102 accreted a gas-rich companion, 
and that much of the gas has now been converted into stars. 
Davidge (2010) finds a concentration of AGB stars to the west 
of the nucleus and suggests that this may be the diffuse remnants of a companion. 

	The central regions of NGC 5102 likely harbor clues into its past evolution, 
as this is the deepest part of the galactic potential and 
is where remnants of accreted stars and material may eventually accumulate. 
If an interaction or some other mechanism triggered large scale radial 
flows of stars and gas then the bulge may also show evidence of recent 
elevated levels of star formation, as well as structural properties that are 
consistent with a disk origin. If the bulge did form in this way then the 
morphology of NGC 5102 has changed with time.

	While studies of resolved stars are one possible means of 
investigating the SFH near the center of NGC 5102, 
the stellar density in the nucleus and bulge is very high. As a result, 
only the brightest stars can be resolved, thereby limiting the age range that can 
be probed. In contrast, spectroscopic studies of integrated light offer a means of 
probing past star-forming activity that is not hindered by crowding.

	In the current study, deep long-slit spectra at 
visible wavelengths are used to probe the stellar content along the minor axis 
of NGC 5102. Well-calibrated spectroscopic diagnostics 
such as the Balmer lines of Hydrogen and absorption features of Mg and Fe near 
5000\AA\ provide information about age, metallicity, and chemical mixture.
Additional clues into the evolution of a galaxy can be 
gleaned from its isophotal properties, and so the photometric and 
structural characteristics of NGC 5102 are also investigated using archival 
SPITZER [3.6] data. Photometric data at $3.6\mu$m are of interest for isophotal 
studies as they sample the red stars that trace stellar mass.

	Details of the observations and the 
processing of the data are provided in Section 2. The isophotal properties of NGC 5102 
are examined in Section 3, and the results are used to define angular intervals 
from which representative spectra of the nucleus, bulge, and disk can be 
extracted. An initial assessment of stellar content is made 
from Lick indices and broad-band colors in Section 4. Different SFHs are explored 
in Section 5, where the spectra are compared with models on a pixel-by-pixel (PBP) 
basis. The paper closes in Section 6 with a discussion and summary of the results. 

\section{OBSERVATIONS \& INITIAL PROCESSING}

	The data were recorded with the Gemini Multi-Object 
Spectrograph (GMOS) on Gemini South during the night of March 18, 2008 
as part of program GS-2008A-Q-72 (PI: Davidge). The sky was clear and the image 
quality entry in the headers indicates that 85\%ile seeing prevailed, 
which notionally corresponds to $\sim 1.1$ arcsec FWHM at 5000\AA.
The detector in GMOS during 2008 was a mosaic of three EEV $2048 
\times 6048$ CCDs. Each $13.5\mu$m square pixel sampled 0.073 arcsec on a 
side. Additional details concerning the design and performance of GMOS have 
been given by (Crampton et al. 2000).

	Spectra were recorded through a 1 arcsec wide slit that was positioned 
along the minor axis of NGC 5102. The light was dispersed with the B1200 grating 
($\lambda_{blaze} =  4630\AA$), and the expected 
resolution of this grating $+$ slit combination is 1872. 
The grating was rotated to deliver a central wavelength of 4750\AA, and 
a $g'$ filter was used to suppress signal from higher spectral orders. 
Twelve 960 sec exposures were recorded, yielding a total observing time of 
3.2 hours. The detector was binned $2 \times 2$ during read-out.

	The initial reduction of the data was done with 
GMOS-specific tasks in the Gemini IRAF package, and this included the subtraction of a 
bias frame and the division by a flat field frame. The latter was obtained 
from dispersed images of the slit as it was illuminated by the 
Gemini facility calibration unit (`GCAL') continuum lamp. Spectra were also 
recorded of the GCAL CuAr lamp immediately after each science exposure to check the 
wavelength calibration and correct for flexure. 
Emission features in the arc spectra have 2.7 \AA\ 
FWHM, which is consistent with the expected spectral resolution.

	The background sky was measured at the ends of the slit 
and the result was subtracted from the wavelength-calibrated 
spectra. The individual wavelength-calibrated exposures were then summed to 
produce a composite two-dimensional spectrum. This summed image was in turn binned 
along the spatial direction in 5 pixel (i.e. 0.73 arcsec) increments 
to boost the S/N ratio in each spatial resolution element.

\section{ISOPHOTAL MEASUREMENTS}

	The isophotal properties of a galaxy are one part of its fossil record, 
and these have been investigated in NGC 5102 at visible wavelengths by Sohn, Chun, 
\& Byun (1992). That study used information gleaned 
from a deep photographic IIa-O image to characterize the 
bulge and disk components. The bulge was found to follow an r$^{1/4}$ light profile and 
to contribute the same amount of light as the disk at a major 
axis distance of $\sim 60$ arcsec.

	The isophotal properties at visible wavelengths can be 
skewed by even modest numbers of young stars, as the brightest members of a 
young population can dominate the light even though they may 
contribute modestly to the total stellar mass (e.g. Serra \& Trager 2007). 
Line emission may also complicate efforts to trace the distribution of stellar mass, 
as it is expected to be concentrated around ionizing sources. 
Given that the central regions of NGC 5102 harbor a substantial population of 
moderately young stars and there is a complex web of line emission (McMillan 
et al. 1994), then it is of interest to conduct an isophotal analysis of 
NGC 5102 in the infrared, where the light better traces total stellar mass. 
For the current study the isophotal properties of NGC 5102 are examined 
using archival SPITZER [3.6] data.

	A Post-Basic Calibrated Dataset (PBCD) mosaic that was constructed from 
[3.6] images that were recorded for program 80072 (PI: Tully) was downloaded from the 
Spitzer IPAC archive \footnote[1]{http://irsa.ipac.caltech.edu/Missions/spitzer.html}.
Isophote properties were measured with the program {\it ellipse} (Jedrzejewski 1987), 
as implemented in the IRAF STSDAS package. The mean surface brightness, the 
ellipticity, and the coefficient of the fourth order cosine term in the Fourier 
expansion of the isophotes -- B4 -- were measured. The latter distinguishes between 
`boxy' and `disky' isophotes (e.g. Carter 1978).

	The isophotal properties of NGC 5102 are shown in Figure 1, where radius 
is measured along the semi-minor axis for consistency with the GMOS spectra. 
The error bars show the uncertainties computed by {\it ellipse}. The x-axis has been 
scaled so that an $r^{1/4}$ law produces a linear trend in the light profile. 

	The bulge is characterised by an $r^{1/4}$ relation and 
dominates the minor axis light profile between 2 and 30 arcsec. 
The dotted line in Figure 1 is an r$^{1/4}$ relation that was fit to the bulge 
light profile -- this relation has an effective surface brightness of $22.464 \pm 
0.004$ mag arcsec${^2}$ and an effective scale length along the minor axis of $25.03 
\pm 0.16$ arcsec. The ellipticity changes with radius in the bulge, with the 
isophotes becoming progressively flatter as radius increases. While B4 has both 
disky (B4 $> 0$) and boxy (B4 $< 0$) values in the 
bulge, disk-shaped isophotes predominate.

	The outer boundary of the nucleus and the inner boundary of the disk are 
identified as the points where the light profile departs from the r$^{1/4}$ relation. 
The nucleus departs from the light profile of the surroundings 
at radii $< 2$ arcsec. Given that the HWHM of [3.6] 
images listed in the IRAC Instrument Handbook \footnote[2]
{http://http://irsa.ipac.caltech.edu/data/SPITZER/docs/irac/iracinstrumenthandbook/5/} 
is 0.84 arcsec, then significant blurring of the structural properties by the 
instrument PSF is expected within the nucleus. Keeping this caveat in mind, the 
ellipticity jumps by $\sim 0.05$ (i.e. the isophotes become flatter) 
near the nucleus/bulge boundary, while the B4 measurements 
are indicative of boxy isophotes in the nucleus, as opposed to 
the disky isophotes that prevail in the bulge.

	The disk dominates the infrared light at radii $> 30$ arcsec, and the change 
in the light profile from an r$^{1/4}$ relation is accompanied by changes in other 
structural characteristics near the bulge/disk boundary. The radial trend 
in ellipticity that prevails throughout the bulge changes near the bulge/disk 
boundary, in the sense that the tendency for the ellipticity to increase with radius 
eventually reverses. The sign of the B4 coefficients also change 
near the disk/bulge boundary, transitioning from disky values in the 
outer portions of the bulge to boxy values at $r > 40$ arcsec. 

\section{LINE STRENGTHS \& GRADIENTS}

\subsection{Spectral Extraction and Index Measurement}

	Spectra that sample different radii were extracted 
from the final two-dimensional spectrum to examine 
radial trends in line strengths, and gauge spectroscopic uniformity 
with radius. The S/N ratio of the two-dimensional spectrum decreases 
with increasing distance from the galaxy center and -- when necessary -- 
the extracted spectra were binned to obtain a S/N ratio $\geq 15$. 
While little or no binning was required to meet this criterion in the nucleus 
and throughout much of the bulge, the extracted spectra in the disk 
include data that spans several arcsec on the sky. 

	The slit passes through an object located 42 arcsec south of the center 
of NGC 5102 along the minor axis. This object has a stellar morphology and a 
velocity that is comparable to NGC 5102, suggesting that it may be a compact star 
cluster. The Lick indices measured for this object differ from 
those in the adjacent disk, and the spatial interval 
containing this object was excluded when extracting spectra.

	An initial reconnaisance of stellar content is conducted 
by examining Lick indices (Worthey et al. 1994). These are well-calibrated 
indices that multiplex information over moderately broad wavelength intervals, 
facilitating the study of absorption features in spectra that may have 
low S/N ratios. While the indices are centered on prominent absorption features 
that originate from one element or molecule, in reality they are 
contaminated by lines from other species (e.g. Worthey et al. 1994), and this 
complicates their use as probes of stellar content. The indices were 
measured using the bandpasses defined on Guy Worthey's website \footnote[3]
{http://astro.wsu.edu/worthey/html/index.table.html}.

	The indices measured in NGC 5102 probe age (H$\beta$) 
and chemical composition (Mg$_2$, Mgb, Fe5270, Fe5335). These indices 
sample some of the strongest absorption features in the visible part of the 
spectra of composite stellar systems, and have comparatively robust 
transformation properties (Puzia et al. 2013). The Fe5270 and Fe5335 indices 
are averaged together for this study to boost the S/N ratio of the Fe features, 
and the mean index is referred to as $<Fe>$. The [MgFe]' index defined by Thomas, 
Maraston, \& Bender (2003), which combines the Mgb, Fe5270, and Fe5335 
indices to track total metallicity in a manner that is not sensitive to 
variations in [$\alpha$/Fe], is also calculated.

	Figure 2 of McMillan et al. (1994) indicates that 
line emission spanning arcmin angular scales -- which corresponds to kpc 
spatial scales at the distance of NGC 5102 -- is present near the center of NGC 5102.
It is thus not surprising that emission features are seen in 
some of the extracted spectra. An index to gauge the strength of the [OIII]5007 line, 
which is a prominent emission feature in many of the extracted spectra, was 
defined. The strength of the [OIII]5007 line is measured in the interval 4995 -- 5020 
\AA , with continua measured in the intervals 4973 -- 4995 \AA\ and 5020 -- 5047 \AA . 
The [OIII] index is measured in \AA , and becomes 
more negative as the emission line increases in strength. 

	The extracted spectra were subject to additional processing prior 
to measuring indices. First, an empirical correction for the wavelength response 
of the instrument and optics was constructed from the galaxy spectra and applied 
to the data. Indices that cover broad wavelength intervals, such as Mg$_2$, can be 
affected by such response functions if the spectrum 
is dominated by stars with cool temperatures. In an effort to minimize the impact 
of the continuum correction on molecular features, the continuum 
was measured from the nuclear spectrum, which has weak metallic features and 
is dominated by deep Balmer absorption lines that can be filtered out when 
obtaining a response function.

	Second, each extracted spectrum was corrected for the bulk radial velocity of 
the galaxy and for differences in velocity due to rotation internal to 
NGC 5102. The results were then smoothed with a Gaussian to match the 
spectral resolution of the Lick system ($\sim 9\AA $). The width of the Gaussian 
varied with location in the galaxy to correct for gradients in velocity dispersion.

	Examples of extracted spectra in the 4500 - 5400\AA\ wavelength interval 
are compared in Figure 2. Spectroscopic features sampled by the 
indices that are examined in this study are indicated. Emission from [OIII]5007 
and H$\beta$ is seen in the $+13$ arcsec spectrum, and so the Lick H$\beta$ index 
measured from this spectrum will likely underestimate the depth of H$\beta$ 
absorption.  While the Lick indices that sample metallic features have passbands that 
avoid prominent emission features like [OIII]5007, they are still subject to 
contamination from continuum emission, which veils absorption features 
and causes their strengths to be underestimated. 

	Puzia et al. (2013) investigate the transformation of GMOS data into the 
Lick system. They consider two instrumental configurations that involve 
a common grating (B600) but that have different resolutions due to pixel binning 
and slit width. The offsets between the instrumental and standard values for 
the majority of indices examined by Puzia et al. (2013) 
differ significantly between the two set-ups. There are 
exceptions, and Puzia et al. (2013) identify H$\beta$, Mgb, Mg$_2$, 
Fe 5270, and Fe 5335 as having `small and well-defined correction terms'. The 
stable nature of the transformations for these indices is probably due to the relative 
strengths of the targeted absorption features, which makes them less susceptible to 
blending from weak contaminating features that fall within the index passbands.

	The linear $\delta$ offsets listed in Table 5 of Puzia et al. (2013) 
apply to the B600 grating with a 0.5 arcsec wide slit and $1 \times 1$ binning. 
This configuration delivers the same resolution as the NGC 5102 GMOS data (i.e. 
the B1200 grating with 1 arcsec slit and $2 \times 2$ binning). Therefore, 
the offsets in Table 5 of Puzia et al. (2013) are used to transform the 
instrumental GMOS indices into the standard system. 

	While the indices considered here have robust transformation properties, 
differences in -- say -- the spectral response of the B600 and 
B1200 gratings might still affect the transformation in subtle ways. 
It is thus encouraging that the various indices predict a consistent metallicity that 
agrees with that measured from resolved RGB stars in the NGC 5102 disk (Section 4.3).
The uncertainties in the transformation notwithstanding, 
the indices obtained here are in an internally consistent system that is 
well-suited for assessing differential changes with radius.

\subsection{Radial Trends}

	The H$\beta$, Mg$_2$, Mgb, $<Fe>$, and [MgFe]' indices 
obtained from the extracted spectra are shown in Figure 3. The comparatively 
sparse angular coverage at large radii reflects the binning applied to 
meet the minimum S/N criterion described in the previous section.
The error bars show the random uncertainties in the indices 
computed from the signal in the passbands -- these 
are smallest near the galaxy center, where the signal is highest. 

\subsubsection{Line Indices: The Nucleus}

	The H$\beta$ index peaks in the central few arcsec of NGC 5102, 
indicating that the nucleus has a younger luminosity-weighted age than the 
surroundings. The region with deep H$\beta$ absorption and the area associated 
with the nucleus in the SPITZER [3.6] image have similar angular extents, suggesting 
that the prominence of the nucleus in the Spitzer data is related -- at least in 
part -- to stellar population factors, and is not driven solely by stellar 
mass concentration. The entries in the bottom panel of Figure 3 suggest that [OIII] 
emission near the nucleus is modest when compared with what is seen at some points 
outside of the nucleus. That the H$\beta$ indices in the nucleus are not skewed 
greatly by H$\beta$ emission is demonstrated in Section 5, where model 
spectra are compared directly with the observations.

	Weak metallic absorption features are detected 
in the nucleus (e.g. Figure 2). The strengths of the H$\beta$ and metallic indices are 
anti-correlated near the nucleus, in the sense that the metallic indices strengthen 
as H$\beta$ weakens with increasing radius. This behaviour is due to the veiling of 
metallic absorption features in the spectra of cool stars by the continuum from 
hot main sequence stars. 

\subsubsection{Line Indices: The Bulge}

	The H$\beta$ index is systematically higher in the interval --5 to --20 
arcsec than in the corresponding range of radii on the other 
side of the nucleus. This may suggest that there are younger stars immediately 
to the south of the nucleus than there are to the north of the nucleus. 
However, the dip in H$\beta$ in the interval 10 -- 15 arcsec is also an 
area of [OIII] emission, raising the possibility that line emission has filled the 
H$\beta$ absorption feature in this angular interval. Thus, the radial age 
properties of the bulge may be more uniform than indicated in Figure 2.

	H$\beta$ and the majority of metallic indices are stable within 
their uncertainties in the bulge when $\mid r \mid > 20$ arcsec. 
While the Mg$_2$ index is asymmetric about the nucleus, 
in the sense that between --20 and --40 arcsec it is a few 
hundredths of a magnitude larger than between 20 and 40 arcsec, this 
may be due to low-level scattered light (see below). In fact, the 
symmetric nature of the [MgFe]' index throughout the bulge suggests that there is 
not a radial metallicity gradient -- in terms of overall metallicity the 
bulge of NGC 5102 appears to be well-mixed. 

\subsubsection{Line Indices: The Disk}

	The Mgb and $<Fe>$ indices have different 
radial behaviours in the disk. While the Mgb index stays constant 
throughout the bulge and disk, the $<Fe>$ index drops at 
the bulge/disk interface. A calculation of mean indices confirms that 
the $<Fe>$ distributions in the disk and bulge differ, while Mgb stays constant. The 
mean $<Fe>$ is $0.97 \pm 0.09\AA $ in the disk and $1.30 \pm 0.02\AA $ 
in the bulge, where the quoted uncertainties are the formal errors in the means. 
The $<Fe>$ point at the southernmost end of the disk, which is by far the highest 
$<Fe>$ measurement in the NGC 5102 dataset, was not 
included when calculating the mean $<Fe>$ for the disk. 
This point is the outlier in the upper right hand corner of the [MgFe]' $vs <Fe>$ 
diagram in Figure 5 (discussed below). For comparison, the means of the Mgb 
indices in the disk and bulge are $2.10 \pm 0.10\AA $ and $2.13 \pm 0.04\AA $, 
respectively. Like Mgb, the averages of the [MgFe]' indices in the bulge and 
disk do not show statistically significant differences, suggesting 
that the overall metallicities of the bulge and disk are similar.

	The Mg$_2$ indices in the southern disk tend to be larger than those on the 
other side of the galaxy, continuing the trend in the bulge noted in the previous 
section. The Mgb indices do not show a north/south asymmetry, and we suspect that the 
high Mg$_2$ indices in the south may be an artifact of uncertainties related 
to continuum shape. The spectra were recorded during moderately 
bright moon conditions, and the behaviour of the Mg$_2$ 
indices may reflect subtle differences in scattered light across the GMOS 
detector that are only evident in indices that span 100 + \AA .

\subsection{Comparisons with Model Indices}

	Indices were measured from model spectra of simple stellar populations (SSPs) 
that were downloaded from the Bag of Stellar Tricks and Isochrones 
(BaSTI) data base \footnote[4]{http://albione.oa-teramo.inaf.it/} (Manzato et al. 
2008; Percival et al. 2009). The models include evolution on the AGB.
SSPs are the basic building blocks for models of more complicated SFHs and 
yield luminosity-weighted effective ages when used on their own.

	The choice of metallicities for the models 
was based on the colors of resolved RGB stars in the disk of NGC 5102. Davidge (2008) 
found that RGB stars in NGC 5102 have [M/H] between --0.9 and --0.1, 
and that the metallicity distribution function peaks near --0.6. Based on these 
results, models with two metallicities were considered: solar and Z=0.004 
(i.e. [M/H] $\sim -0.7$). A solar chemical mixture was adopted 
as there is no information on the chemical enrichment history of NGC 5102.
If the chemical mixture in NGC 5102 differs from solar then offsets 
between the observed and modelled Mgb and $<Fe>$ indices might be expected. 

	The model spectra were processed to duplicate the steps applied to the 
extracted GMOS spectra. In particular, the models were re-sampled 
to match the dispersion of the GMOS data, and then 
continuum-corrected by applying a fit that was obtained from a young population 
using the same fitting parameters (i.e. the same function 
type, order, and rejection limits) that were applied to the GMOS data. Finally, 
the model spectra were smoothed to match the spectral resolution of the Lick system.

	The observed and model indices are compared 
in Figure 4. It is evident that (1) the central regions of the nucleus 
have a luminosity-weighted age at visible wavelengths of a few hundred Myr, 
and (2) the disk and bulge having substantially older ages than the nucleus. 
The luminosity-weighted age of the nucleus increases with radius within 
the nucleus, and this may be due in part to seeing. In addition, that the innermost 
few parsecs of NGC 5102 may harbor a population with an even younger luminosity-weighted 
age than is found here can not be ruled out given the limited 
angular resolution of these data. The effective age of 
the nucleus deduced from the indices is not greatly different from that found 
by Miner et al. (2011), who sampled wavelengths where the light 
originates from a different mix of stellar types than at 
visible wavelengths. That the nucleus contains a concentration of 
stars with ages in excess of a few hundred Myr provides a natural 
explanation for the prominent nature of the nucleus in the [3.6] SPITZER 
images, as a large population of luminous AGB stars are expected in such a system 
(Section 4.4).

	With the exception of the measurements on the H$\beta\ vs$ Mg$_2$ plane, the 
observed indices throughout the nucleus and the inner regions of 
the bulge are consistent with a sub-solar metallicity, with 
the nucleus and inner bulge indices intersecting or falling close to the Z=0.004 
sequences. In the previous section it was demonstrated that the Mgb and $<Fe>$ 
indices in the bulge and disk have different distributions. 
This is also evident in Figure 4, where the bulge and disk points on the 
H$\beta\ vs$ Mgb diagram overlap on the Mgb axis, while there is a clear tendency for 
the disk points in the H$\beta\ vs <Fe>$ diagram to fall to the left of the 
bulge measurements. 

	The H$\beta$ indices throughout much of the disk are smaller than predicted 
by the 10 Gyr models. The fitting of model spectra 
to the observations in Section 5 reveals that H$\beta$ emission may 
partially fill in H$\beta$ absorption, causing points in the various panels of 
Figure 4 to fall below where they would if there was no emission.
Given that H$\beta$ can be affected by line emission, it is then of 
interest to make comparisons with model sequences using only metallic indices. 

	Comparisons between the observations and models that involve only metallic 
features are made in Figure 5. The overall trends defined by the Z=0.004 and solar 
models are not markedly different in this figure. The modest offsets between the two 
model sequences in the various panels are due in part to differences in the temperature 
of the main sequence and red giant branch, and the resulting effect on the depth of 
metallic features. 

	The NGC 5102 Mg$_2$ indices are clearly offset from the 
model indices on the [MgFe]' $vs$ Mg$_2$ plane, although 
the trends defined by the observations and models parallel each other. 
The application of an offset of a few hundredths of a magnitude  
to Mg$_2$ will force agreement between the models and observations in the 
left hand panel of Figure 5. This same correction would also produce agreement between 
the observations and Z=0.004 models in the upper left hand panel of Figure 4.

	The solar metallicity models extend to much higher index values than 
the Z=0.004 models in Figure 5, and the full extents of the solar metallicity sequences 
are not shown in this figure. It is thus significant that 
the range of indices in NGC 5102 matches that predicted 
by the Z=0.004 models. Moreover, the NGC 5102 data fall within a few tenths of 
an \AA\ of the Z=0.004 models on the H$\beta\ vs$ Mgb and H$\beta\ vs <Fe>$ planes. 
Whereas the majority of points fall to the left of the Z=0.004 sequence 
on the [MgFe]' $vs$ Mgb diagram, the majority of points on the [MgFe]' $vs <Fe>$ 
diagrams fall to the left of the model sequence. This again signals that 
the chemical mixture in NGC 5102 likely differs from solar.

\subsection{Broad-Band Colors}

	Broad-band photometry provides a check on the stellar content information 
that is obtained from spectra. In general, leverage for probing stellar 
content increase when the wavelength range is broadened, and so 
photometry that covers a wide range of wavelengths is prefered.
For the current study, colors are measured from visible and mid-infrared images.

	Photometric measurements at visible wavelengths were obtained from 
the 120 sec $g'$ acquisition exposure that was recorded 
immediately before the GMOS spectra. As per Gemini acquisition procedures, only 
the central GMOS CCD was read out. The data were 
flat-fielded using a sky flat that was constructed from other GMOS $g'$ acquisition 
images that were recorded during March 2008. Photometry that sampled longer 
wavelengths was obtained from the [3.6] SPITZER image discussed in Section 3, and the 
[4.5] SPITZER image obtained from data recorded for the same program.
The [3.6] and [4.5] observations have an angular resolution that is 
not greatly different from that of the GMOS data, simplifying the comparison 
between colors and line indices.

	The SPITZER images were transformed to match the pixel scale and orientation 
of the $g'$ image. The $g'$ and [3.6] images were then smoothed to match the 
angular resolution of the [4.5] data, which has the poorest angular resolution 
of the image trio. All three images were sky-subtracted using measurements 
along the minor axis made near the edges of the fields.

	The $g'-[3.6]$ and [3.6]--[4.5] colors are shown in Figure 6. 
The radial coverage is restricted to $\pm 25$ arcsec about 
the nucleus as the S/N of the SPITZER data plunges at larger radii. 
The SPITZER observations were calibrated using zeropoints from Reach et al. 
(2005), while the $g'$ measurements were calibrated using archived photometric 
zeropoints. Jorgensen (2009) investigates the time variation of GMOS zeropoints, and 
these variations suggest that the use of archived zeropoints introduces an 
uncertainty of $\pm 0.1$ magnitudes into the $g'-[3.6]$ calibration.

	The H$\beta$ measurements from Figure 3 are re-plotted in the top panel of 
Figure 6 to assist in the interpretation of the results. The mix 
of stellar types that contributes to the integrated light changes with wavelength, 
and so the amplitude of any color variations depends on the wavelength 
range that is examined. In the absence of emission and absorption by 
dust at visible wavelengths the nucleus of NGC 5102 should have a bluer 
color than the bulge, and this is observed to be the case (Pritchet 1979). 
However, the nucleus in $g'-[3.6]$ is redder than the surroundings. This is due 
to luminous AGB stars that dominate the infrared light in the nucleus. 
Evidence for a significant AGB component is seen in the near-infrared spectrum 
presented by Miner et al. (2011).

	Additional evidence for a nuclear concentration of massive AGB stars is found 
at longer wavelengths. Images from the Wide-field Infrared Survey Explorer (WISE: 
Wright et al. 2010), downloaded from the NASA/IPAC Infrared Science Archive 
\footnote[5]{http://irsa.ipac.caltech.edu/Missions/wise.html}, reveal 
pronounced emission from the central regions of NGC 5102 at $12\mu$m and 
$22\mu$m. The emission at $22\mu$m is of particular interest as it has a compact 
morphology. The WISE images thus reveal that the nucleus of NGC 5102 harbors 
significant quantities of warm dust. However, the blue color of the nucleus at visible 
wavelengths suggests that large quantities of dust likely 
do not pervade the interstellar medium throughout the nucleus. 
Rather, the warm dust detected in the WISE images is probably concentrated 
in circumstellar envelopes that gird individual AGB stars.

	Models of the integrated photometric properties of SSPs from Marigo et al. 
(2008) were downloaded from the Padova observatory website 
\footnote[6]{http://stev.oapd.inaf.it/cgi-bin/cmd}. 
The models assume Z = 0.004, a Chabrier (2001) mass function, 
and an 85\% AMC $+ 15\%$ silicate model for circumstellar dust 
(Groenewegen 2006). This particular circumstellar composition was adopted 
because of the spectroscopic evidence for bright C stars found by Miner et al. (2011). 
However, the dust composition is expected to be a major factor 
only at longer wavelengths than those considered here (e.g. Davidge 2014). 

	The comparisons in Figure 4 indicate that the nucleus has a 
luminosity-weighted age of a few hundred Myr, while the bulge has an effective age 
of a few Gyr. The models predict that a population with an age $> 2$ Gyr will have 
$g'-[3.6] \sim 2.4 - 2.5$, and this is consistent with the $g'-[3.6]$ color 
of the bulge in Figure 6. The models also predict that a population 
with an age 0.5 -- 1.0 Gyr, like that expected in the 
nucleus of NGC 5102, will have $g'-[3.6]$ colors that are a few tenths 
of a magnitude redder than the bulge, and this is consistent with 
the relative $g'-[3.6]$ colors of the bulge and nucleus. Models that are 
generated without an AGB component do not have a red nucleus.

	While less sensitive to age than $g'-[3.6]$, 
the [3.6]-[4.5] color, shown in the bottom panel of Figure 6, 
provides supplemental information about the stellar content of NGC 5102. 
Models predict that [3.6]--[4.5] should be stable to within $\pm 0.01$ magnitude 
for ages $> 2$ Gyr. The [3.6]--[4.5] color between 5 and 15 arcsec 
varies by less than this amount, as expected if the bulge has a uniform age.

\section{PIXEL-BY-PIXEL COMPARISONS WITH MODEL SPECTRA}

	The Lick indices combine signal over many 
\AA , and the resulting decrease in spectral resolution 
can introduce ambiguities in age and metallicity measurements. In contrast,
direct PBP comparisons with model spectra provide a means 
of probing stellar content without compromising spectral resolution. 
In this section PBP comparisons are made with models that are based on two SFH 
families that might be expected in NGC 5102. 

\subsection{Models and Matching Procedures}

	One family of models assumes SSPs. These models 
might be expected to represent the visible spectrum if $\sim 10\%$ or more of 
the stellar mass formed within the past few Gyr (Serra \& Trager 2007). We note that 
the mean star burst amplitude in the sample of dwarf galaxies studied by 
McQuinn et al. (2010) is $<b> \sim 6$, and so large bursts that can be approximated 
as SSPs might be expected. The SSP models assume a Chabrier (2001) mass function.

	The SFHs compiled by Weisz et al. (2011) indicate that if there have been 
multiple star-forming bursts then the SFHs of gas-rich dwarf galaxies can be 
approximated by a constant star formation rate (cSFR) when averaged over suitably 
long timespans. The stellar mass of NGC 5102 is comparable to that of the LMC (Davidge 
2010), and galaxies of this size may experience a higher number of 
star bursts than more massive systems (e.g. Huang et al. 2013). Therefore, a second 
family of models that assume a cSFR was also 
explored. Star formation was assumed to start 12 Gyr ago, and the models were 
constructed by combining SSP models. A Chabrier (2001) mass function 
was adopted. Given the evidence that star-forming 
activity in NGC 5102 plummeted during recent or intermediate epochs, 
then the cSFR models also assume that star formation 
was shut off in the past and was not re-started. 
With the metallicity, SFR, and the IMF set then the sole free parameter of 
these `truncated' cSFR models is the termination time of star-forming activity. 

	The models were constructed from the same family of BaSTI spectra that 
were used in Section 4. In light of the comparisons with the Lick indices, only 
models with Z=0.004 are considered for the PBP comparisons. The use of 
solar metallicity models would result in ages that are $\sim 0.3$ dex 
younger than those found with Z=0.004.

	As was done in Section 4, the model spectra were convolved with a gaussian to 
match the resolution of the GMOS spectra, which is set here by the instrument 
configuration and the central velocity dispersion of NGC 5102 (Section 4). The 
model spectra were also continuum-corrected. Given the problems 
in matching the observed and modelled Mg$_2$ indices 
discussed in Section 4, coupled with the suspected influence of 
scattered light on the spectra extracted for the southern disk, a high order continuum 
function was fit to each spectrum individually, rather than basing the continuum 
on only one spectrum that has weak metallic features. 
While the removal of a high-order continuum function 
alters the depths of broad molecular absorption bands, 
the age estimates should not be biased if the continua in the 
model and observed spectra are measured and removed in a consistent manner.

	Spectra of the nucleus, bulge, and disk of NGC 5102 were 
constructed by co-adding all extracted spectra in these regions. While 
sacrificing angular information, the combined spectra have high S/N ratios. 
Given the modest angular extent of the nucleus and the blurring introduced 
by seeing then the characteristic ages found for the nucleus are expected to 
be upper limits. The co-added spectra were re-sampled to match the 1 \AA\ spacing 
of the BaSTI models, and the results are shown in Figure 7.

	SSP and cSFR models that minimized the residuals 
in the wavelength interval 4750 - 5400 \AA\ were identified. 
This wavelength interval was selected as it has the highest S/N ratio. Still, the 
results do not change significantly if the comparisons are made in the 4100 - 5400\AA\ 
interval -- the key criterion is the inclusion of a Balmer line (H$\beta$ in 
this case) to provide leverage for age estimates.

	Line and continuum emission can skew age estimates, and so 
residuals were measured after supressing points that exceeded the initial dispersion by 
more than the $2.5\sigma$ level. The application of this sigma-clipping criterion did 
not affect the age estimates of the nucleus and disk. However, there 
are prominent emission lines in the bulge spectrum (see below), and sigma-clipping 
resulted in a slightly lower age estimate than when no clipping was applied.

	The uncertainties in the residuals 
were estimated by adding noise to the best-fitting model spectrum using the IRAF 
{\it mknoise} task to simulate the S/N ratio of the spectrum being modelled. 
The dispersion in the age estimate could then be found after running a number of such 
realizations and comparing the results. The uncertainties found in this way take 
into account random noise, but do not include systematic errors introduced by 
mean metallicity or chemical composition. Thus, the uncertainties are lower limits.

\subsection{Results: SSP models}

	The effective ages of the SSP models that best match the observations are 
listed in Table 1. The nucleus is significantly younger than either the bulge or disk, 
while the bulge is in turn significantly younger than the disk.
There is thus a radial gradient in the effective ages obtained from the SSP models, 
in the sense of progressively younger ages towards smaller radii. 

	The residuals that remain after subtracting the best-matching SSP models from 
the nucleus and bulge spectra are shown in Figure 8. [OIII] 4959 and [OIII] 
5007 lines are seen in the bulge and nucleus residuals, and H$\beta$ emission is 
also seen in the bulge residuals. The relative strengths of the [OIII] lines 
provide a check on the agreement between the observations and models, as the 
[OIII] 4959 line should have a strength that is $\sim 0.3\times$ 
that of [OIII] 5007 (Osterbrock 1989). The line strengths in the lower panel of 
Figure 8 are consistent with the predicted ratio.
Finally, while the model fits were restricted to the 
4750 - 5400\AA\ interval, the residuals near H$\gamma$ (not shown) are similar to those 
near H$\beta$, even though line emission is expected to be less of an issue 
for H$\gamma$. 

\subsection{Results: Constant SFRs models}

	For the present work, cSFR models in which star formation 
starts 12 Gyr ago and terminates at 0.2, 0.4, 0.7, 
0.9, 1.3, 1.8, 2.5, 4, 6, 8, and 9.5 Gyr in the past are 
considered. While there is evidence for low-level star-forming activity 
at the present epoch in NGC 5102 (e.g. Section 1), the integrated spectra indicate that
this activity does not contribute significantly to the visible light. Therefore, 
cSFR models in which star formation terminates within the past 0.2 Gyr are not 
considered.

	The truncated cSFR model that provides the best match to the nucleus spectrum 
yields residuals that are larger than those of the best-matching SSP model at the 
many tens of $\sigma$ level. The vastly superior agreement with an SSP model suggests 
that the visible light from the nucleus is dominated by stars that formed during a 
large-scale star-forming event that occured over a 
time interval of no more than a few hundred Myr. However, the situation 
is very different in the bulge. The cSFR model in which star formation 
is truncated 0.7 Gyr in the past produces a match to the bulge spectra 
that is $5\sigma$ better than the best-matching SSP model. 
The bulge of NGC 5102 thus contains a mix of stars that span 
a range of ages, with no single age group dominating. Given that star formation 
in the bulge proceeded up to $\sim 0.7$ Gyr in the past, then there was 
significant star-forming activity in the bulge at the same time that 
there was large scale star-forming activity in the nucleus.

	As for the disk, a cSFR model in which star formation terminated 9.5 Gyr in 
the past (i.e. star formation continued for $\sim 2.5$ Gyr after starting 
12 Gyr ago), gives a better -- but not statistically different -- match to the 
disk spectrum than the SSP model. The comparisons between the GMOS disk spectrum 
and models thus suggest, but do not conclusively prove, that the disk 
of NGC 5102 contains stars that formed over a timespan 
of a few Gyr, and not during a single large burst. The duration of 
star formation in the disk aside, the visible light from the 
NGC 5102 disk is dominated by stars that are -- on average -- much older than 
those that dominate the same wavelength region in the bulge and nucleus. As 
discussed in the next section, this old age is a key element in understanding the 
history of NGC 5102.

\section{DISCUSSION \& SUMMARY}

	Deep long-slit spectra that cover the blue/visible wavelength region 
are used to probe the SFHs of the nucleus, bulge, and inner disk 
of the lenticular galaxy NGC 5102. The wavelength interval that is sampled contains a 
number of prominent atomic and molecular transitions that are traditional probes 
of stellar content. Absorption features are traced out to $\sim 1$ arcmin from the 
galaxy center along the minor axis, which corresponds to a projected linear 
distance of $\sim 0.9$ kpc using the distance modulus measured by Davidge (2008). 
Archival SPITZER [3.6] images are also used to investigate 
the structural properties of NGC 5102.

	Two approaches are used to probe the stellar content. First, Lick indices 
are measured and compared with indices obtained from model 
spectra that have been processed to duplicate the properties of the 
observed spectra. The indices selected for this work sample strong absorption 
features and have robust transformation characteristics. Second, PBP comparisons 
are made with model spectra that assume SSPs and truncated versions of a cSFR. 
In Section 6.1 it is shown that the indices and PBP comparisons reveal a 
radial chronology for star formation in NGC 5102 that differs from what would 
be expected if its present-day properties were shaped by a major merger.

\subsection{Star Formation in the Nucleus, Bulge, and Disk}

	The Lick indices indicate that the visible light from the nucleus is 
dominated by stars that have roughly the same metallicity as the stars that 
dominate the visible light in the bulge and disk. The indices are consistent with 
Z$\sim 0.004$, which is comparable to the mean found for disk RGB stars by 
Davidge (2008). That the metallicity obtained from integrated light matches 
that determined from resolved stars is an important consistency check, and 
suggests that mixing has occured throughout the galaxy.

	Comparisons with models indicate that the 
luminosity-weighted age of the nucleus at visible wavelengths is $\leq 1$ Gyr. 
The nuclear spectrum is better matched by models that assume a SSP 
instead of those that assume long-term star-forming activity with a constant SFR. 
The superior agreement between the nuclear spectrum and an SSP model 
suggests that the visible light from the nucleus is dominated by stars that 
formed over a timespan of no more than a few hundred Myr. 

	That substantial nuclear star-forming activity occured $\sim 1$ Gyr in the 
past is consistent with the relatively red $V-[3.6]$ color of the nucleus, 
which is attributed here to a large population of luminous AGB stars. A large AGB 
component can also explain the emission detected by WISE at $12\mu$ and $22\mu$m. 
Direct evidence of a large AGB component is seen in the near-infrared absorption 
spectrum (Miner et al. 2011). The age and metallicity of the nucleus measured from 
the GMOS spectrum are such that a large C star population might be expected (Maraston 
2005), and Miner et al. (2011) find spectroscopic evidence of such a component. 

	While low levels of nuclear star formation may have continued to very 
recent epochs, the SFR during recent times has been comparatively low. Indeed,
the ultraviolet SED of the center of NGC 5102 is consistent with only a modest 
contribution from main sequence stars younger than 0.3 Gyr (Kraft et al. 2005). 
Still, there are hints that star-forming activity might 
be experiencing a revival near the center of NGC 5102. 
DeHarveng et al. (1997) identify bright stars near the nucleus that may 
be as young as 15 Myr, although they note that these could also be 
older post asymptotic giant branch objects. 

	Additional evidence of recent central star-forming activity comes from 
Beaulieu et al. (2010), who investigate the SFHs of three fields along the 
major axis of NGC 5102. Only the brightest stars in each field are resolved 
due to crowding, and this limits investigations of SFHs 
to the past few hundred Myr. Field F1 studied 
by Beaulieu et al. (2010) samples the central regions of the galaxy, and the 
SFH constructed from their data shows a more-or-less flat SFR 
40 -- 200 Myr in the past, but a five-fold increase in the SFR during the past 20 Myr.

	The level of star-forming activity during the past $\sim 10$ Myr could be 
estimated from the strengths of emission lines. However, the GMOS spectra are not 
well-suited to this task. The GMOS slit samples only a tiny part of the galaxy, 
whereas the line emission in the central regions of NGC 5102 has a complex spatial 
distribution (McMillan et al. 1994) -- this makes it difficult to estimate a 
total SFR from only a single slit pointing. Some of the line emission
may also not be related to recent star formation. Finally, the GMOS spectra 
do not sample H$\alpha$, which is the primary emission line SFR diagnostic 
at visible wavelengths. SFRs estimated from H$\beta$ are subject to significant 
uncertainties due to extinction, while the strengths of the [OIII] lines 
depend on metallicity and the ionization parameter, with the result that they do 
not correlate as well with SFR as other indicators (Moustakas et al. 2006).

	What could cause a recent uptick in central star-forming 
activity? The frequency of blue nuclei among nearby galaxies suggests 
that star formation near the centers of these systems has a 
duty cycle of $\leq 1$ Gyr (Davidge \& Courteau 2002). We speculate that the 
youngest stars near the center of NGC 5102 found by Deharveng et al. (1997) and 
Beaulieu et al. (2010) may be the result of such cyclical star-forming activity. 

	The nucleus of NGC 5102 undoubtedly contains stars that are significantly 
older than those that formed during the dominant burst. Kraft et al. (2005) model the 
SED of the central regions of NGC 5102 as a combination of two components, consisting 
of a young population with (Z/Z$_{\odot})$ = 0.2 (i.e. Z$\sim 0.003$), 
and a dominant older population with (Z/Z$_{\odot}$) = 1.5. The GMOS 
spectra do not show clear evidence of an older, metal-rich component.

	The bulge dominates the light along the minor axis between 3 and 35 arcsec 
in the [3.6] images. The luminosity-weighted age of the bulge at visible 
wavelengths obtained from the GMOS spectrum is $\sim 2$ Gyr. However, in Section 5 
it is shown that the model in which there was continuous star formation up to $\sim 
0.7$ Gyr in the past better matches the bulge spectrum than the SSP model. 
It thus appears that the bulge contains stars that formed in significant 
numbers over a range of epochs. Even though star-forming activity in the bulge 
overlapped in time with the dominant burst in the nucleus, there is no evidence for a 
large amplitude star burst in the bulge. 

	Beaulieu et al. (2010) conclude that 2\% of the bulge mass formed 
within the past 200 Myr, and suggest that much of the bulge may have formed 
during the past Gyr. In fact, while the bulge of NGC 5102 undoubtedly contains 
stars that are younger than 0.7 Gyr, the GMOS spectra suggest 
that they are not the dominant contributor to the visible light.
Indeed, if much of the bulge formed during the past 1 Gyr then there would be 
much deeper Balmer absorption lines than are seen in the bulge spectrum. 

	A prediction based on the current work is that higher angular resolution 
observations of the NGC 5102 bulge will find evidence of 
SFRs many Gyr in the past that are substantially higher than those during the past 
few hundred Myr. One way to probe directly the SFH of such a crowded environment would 
be to observe in the mid-infrared with a space-based telescope 
having an aperture of a few meters or larger. This 
is a wavelength range where bright AGB stars that span 
a range of ages stand out against the main body of the much fainter 
and bluer objects that dominate the stellar mass. The luminosity 
function of AGB stars over a modest range of mid-infrared magnitudes samples objects 
that formed over a broad range of epochs (e.g. Davidge 2014). 

	The disk dominates the light along the minor axis at radii $> 35$ arcsec, 
and has a luminosity-weighted age $\sim 10$ Gyr if the metallicity is 
Z=0.004. Comparisons with cSFR models in which star formation proceeded for only 
2 -- 3 Gyr yield slightly better agreement with the disk spectrum than SSP models. 
The duration of star-forming activity notwithstanding, large-scale star formation 
in the disk of NGC 5102 evidently shut down many Gyr in the past, well before 
the drop in SFR in the nucleus and bulge. The disk is not completely 
devoid of recent star-forming activity, as there is a diffuse population of 
bright blue stars (Davidge 2010). 

	An old luminosity-weighted age for the disk of NGC 5102 is at odds with the 
conclusion reached by Davidge (2008; 2010) that the large number of AGB stars found 
in the disk belong to a dominant population with an age of a few Gyr. However, 
as demonstrated by Davidge (2014), photometric variabity among AGB stars should be 
considered when estimating ages, and failure to do so will result in age estimates 
that are biased to significantly younger values. Davidge (2008) did not account for 
photometric variability, and so likely underestimated the ages of AGB stars.

\subsection{The Progenitor of NGC 5102}

	Given that a significant fraction of the bulge population in NGC 5102 does 
not have a primordial origin, but formed a few Gyr in the past, then the 
bulge-to-disk ratio of NGC 5102 at the present day almost certainly differs from 
what prevailed 5 -- 10 Gyr in the past. Thus, the morphology of NGC 5102 has 
likely changed during the past few Gyr. The past morphology of 
NGC 5102 is of interest as it provides clues into (1) the types of galaxies 
that might eventually evolve into lenticular systems in low density environments 
like the Centaurus Group, and (2) the mechanisms that 
cause such a transformation. Two possible early morphologies 
for NGC 5102 are considered here. The first 
is a late type spiral galaxy. Such a system might 
transform into a lenticular galaxy if it experienced a merger, as suggested 
by Davidge (2008; 2010) and Beaulieu et al. (2010) for NGC 5102.

	Interactions and mergers have the potential 
to trigger galaxy-wide star-forming activity and alter 
morphology, and so play a key role in galaxy evolution. Tidal interactions 
during such an event can form a bar that channels gas into the central regions 
of the primary galaxy (e.g. Hopkins et al. 2009). The area of active star formation 
shrinks in size as gas is depleted and/or is removed from the outer regions of a disk
(e.g. Soto \& Martin 2010), and the inner regions of the galaxy are expected to contain 
the last remnants of elevated levels of star formation. 
A pseudo-bulge might form when the bar buckles, and this 
structure will contain a mix of virialized stars from the progenitor 
and stars that formed after the merger. If the gas channeled into the bar 
originated in the disk of the progenitor then it 
would have a disk-like chemical mixture and structure. The isophotal 
analysis in Section 3 indicates that the bulge of NGC 5102 has a disky shape.

	The difference between the effective ages of the bulge and nucleus of 
NGC 5102 is comparable to the damping time for starbursts. 
Indeed, the timescales of bursts in nearby dwarf galaxies range from 0.45 to 
1.3 Gyr (McQuinn et al. 2010). However, a merger fails to explain the old age of 
the NGC 5102 disk. Indeed, if the dominant burst of star formation in the nucleus was 
triggered by a merger, then the disk might be expected to show evidence of 
large scale star formation $\sim 1 - 2$ Gyr in the past, and this is not the 
case in NGC 5102. In addition, depending on the geometry of the interaction 
and the mass ratio of the interacting galaxies then the existing stellar disk can be 
completely obliterated in such an encounter (e.g. Hopkins et 
al. 2009), whereas a very old disk is in place in NGC 5102. 
Finally, it might be expected that the bulge would contain a large 
fraction of stars that formed during and immediately after the interaction, but this 
is not consistent with the best-matching cSFR model, which suggests that there 
was not a recent large scale burst like that in the nucleus.

	The problems with transforming a late-type spiral into a 
galaxy with stellar content like that in NGC 5102 motivates the consideration 
of a second progenitor morphology, in which NGC 5102 was originally a barred disk 
galaxy. A key element of this model is that the 
bar was a long-lived structure, in contrast with a short-lived 
tidally-induced structure. The bar buckled, possibly due to the formation of 
a central mass concentration as gas was channeled into the central regions (e.g. 
review by Kormendy \& Kennicutt 2004), resulting in the present-day lenticular galaxy. 
It should be emphasized that NGC 5102 may very well have interacted with another 
galaxy -- and there are hints of tidal interactions (Section 1) -- 
but the point to be made here is that the primary 
driver of the present-day properties of NGC 5102 was a long-lived bar.

	An intriguing possibility is that NGC 5102 may have been a barred dwarf 
irregular galaxy like the LMC. In fact, even though interactions with the SMC 
and the Galaxy have shaped star-forming events in the LMC, the stellar 
contents of NGC 5102 and the LMC are broadly similar. Of particular 
note is that the disk of the LMC appears to have experienced outside-in 
growth (e.g. Meschin et al. 2014; Piatti \& Geisler 2013). Harris \& 
Zaritsky (2009) map the SFH of the LMC, and find significant star-forming activity in 
the outer disk at early epochs. However, by intermediate epochs the bulk of the 
star-forming activity has shifted to the bar, where it persists to the present day. 
If the bar in the LMC were to buckle then the result would be a bulge that contains 
stars spanning a range of ages, surrounded by an old disk.

	A SFH in which there is an old disk that experienced a drop in 
star formation early-on, coupled with a long-lived star-forming bar, could 
explain the differences in the chemical mixtures found in Section 4. 
These differences suggest that the stars in the disk of NGC 5102 
formed from gas that experienced more rapid enrichment (i.e. have higher [Mg/Fe]) than 
the gas from which the majority of bulge stars formed. This is counter to the 
conventional enrichment picture in which stars in classical bulges in early-type 
spirals show evidence of forming from gas that experienced rapid chemical enrichment, 
whereas disk stars form from gas that experienced a more leisurely enrichment 
history. If chemical enrichment in the NGC5102 disk was effectively truncated 
early-on and was restricted to a short timespan (e.g. $\leq 1$ Gyr) then an Fe 
deficiency might be expected. In contrast, if the stars in the bulge formed over an 
extended period of time then they will not be as Fe-deficient as those in the disk.

	We close by noting that the evolution of NGC 5102 may provide clues into the 
origins of other lenticular galaxies that are not in crowded galaxy clusters. 
NGC 404 is a dwarf S0 galaxy at a distance that is comparable 
to NGC 5102, and that has a stellar content that is reminiscent of 
NGC 5102. The resolved stellar content of the disk of NGC 404 indicates that it is 
dominated by old stars (Williams et al. 2010), and long-slit spectra reveal 
a centrally concentrated young and intermediate-age component (Bouchard et al. 2010). 
While NGC 404 shows signs of recent merger activity (e.g. Bouchard et al. 2010), 
such an event did not result in the transfer of large 
quantities of gas to NGC 404 (Bresolin 2013). In fact, as with 
NGC 5102, the age difference between the disk and central regions 
of NGC 404 suggests that a merger may not have been the mechanism that defined 
the present-day appearance of that galaxy.

\acknowledgements{Thanks are extended to the anonymous referee for pointing out the 
pronounced WISE Band 3 and Band 4 emission from the nucleus of NGC 5102.}

\parindent=0.0cm

\clearpage

\begin{table*}
\begin{center}
\begin{tabular}{cc}
\tableline\tableline
COMPONENT & AGE (Gyr) \\
\tableline
Nucleus & $1.0^{+0.2}_{-0.1}$ \\
Bulge & $2.0^{+0.5}_{-0.2}$ \\
Disk & $10^{+2}_{-2}$ \\
\tableline
\end{tabular}
\caption{SSP AGES FROM PIXEL FITTING}
\end{center}
\end{table*}

\clearpage

\begin{figure}
\figurenum{1}
\epsscale{1.00}
\plotone{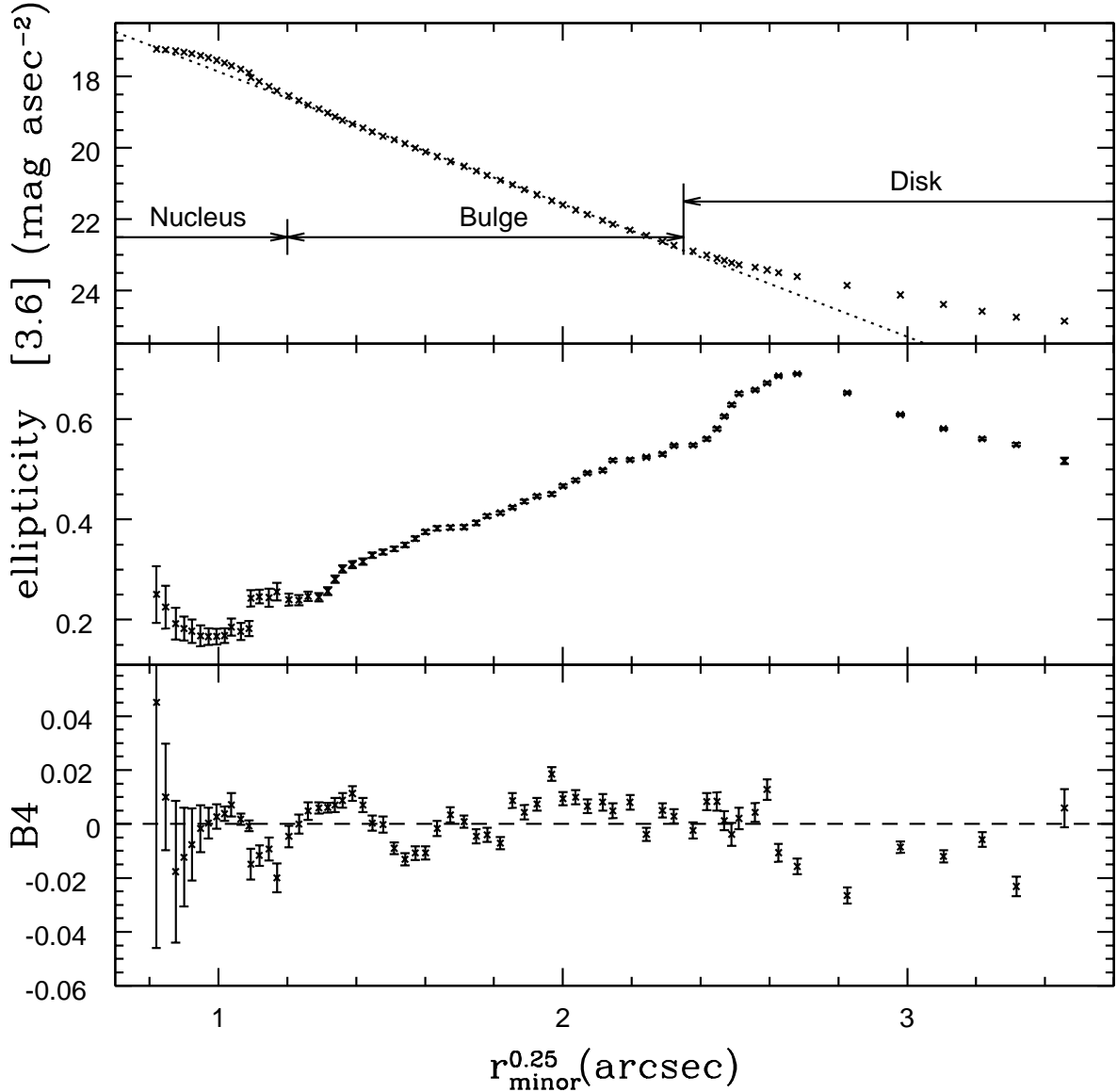}
\caption{Isophotal properties of NGC 5102 in [3.6]. The radial runs of 
surface brightness, ellipticity, and B4 -- the fourth-order coefficient of 
the cosine term in the Fourier representation of the isophotes -- are shown. 
The error bars show the uncertainties computed by {\it ellipse}. 
r$_{minor}$ is the radius in arcsec measured along the minor axis. The dotted line 
in the top panel is an r$^{1/4}$ law that was fit to the bulge light profile. The 
points in the middle and bottom panels indicate that the nucleus/bulge and bulge/disk 
transitions are accompanied by changes in isophote shape. The dashed line in the 
bottom panel marks B4 = 0. Whereas the isophotes throughout much of the bulge have a 
`disky' shape, the isophotes in the nucleus and disk tend to be `boxy'.}
\end{figure}

\clearpage

\begin{figure}
\figurenum{2}
\epsscale{1.00}
\plotone{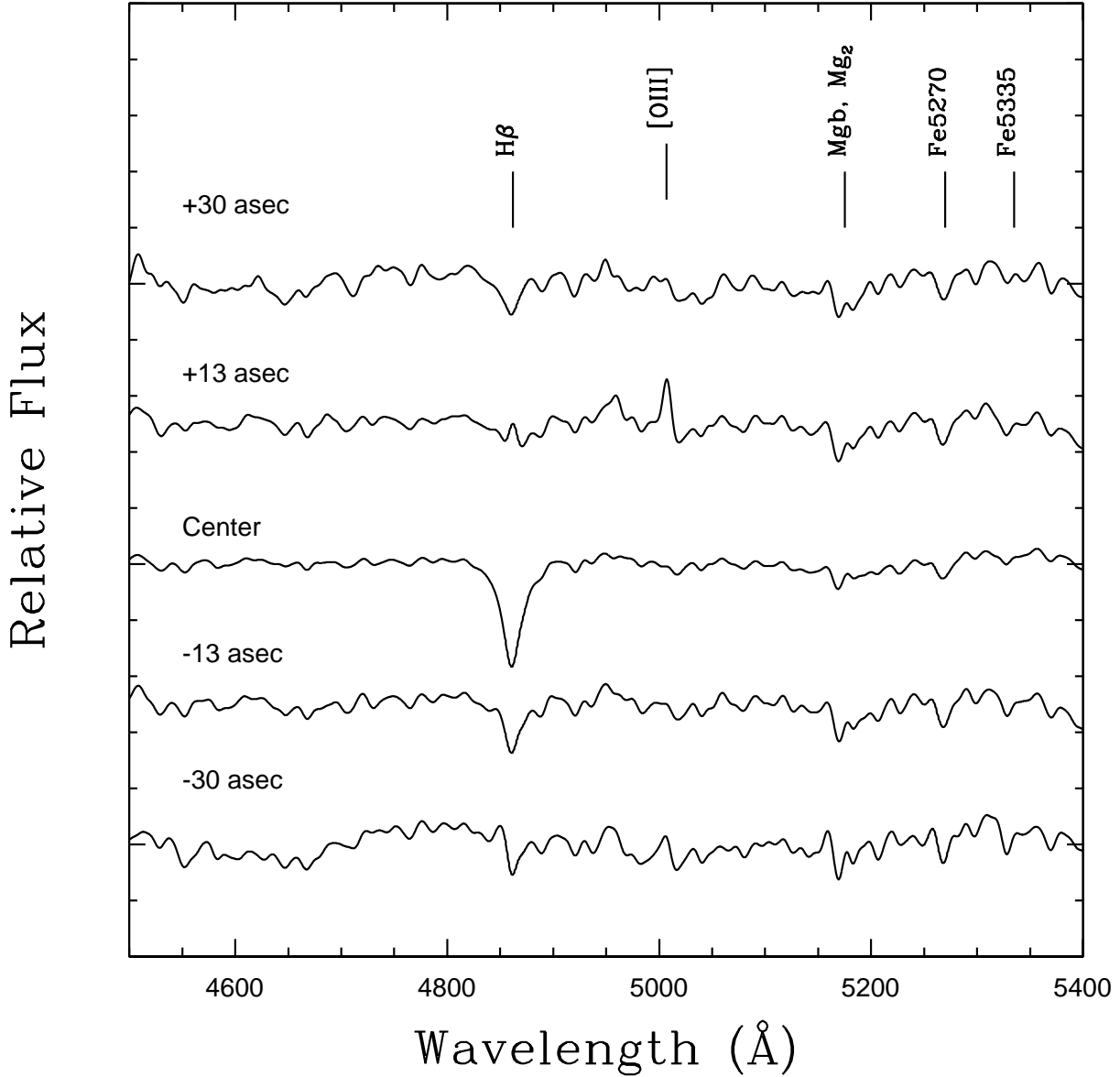}
\caption{Sample extracted spectra. The spectra were produced using the procedure 
described in the text. The spectra shown here have been normalized to unity and then 
shifted along the vertical axis to facilitate comparison. Spectral features that are 
associated with the indices considered in this study 
are flagged. Radial offsets along the minor axis from the galaxy 
center are indicated, with positive values to the north of the galaxy 
center, and negative values to the south.}
\end{figure}

\clearpage

\begin{figure}
\figurenum{3}
\epsscale{0.80}
\plotone{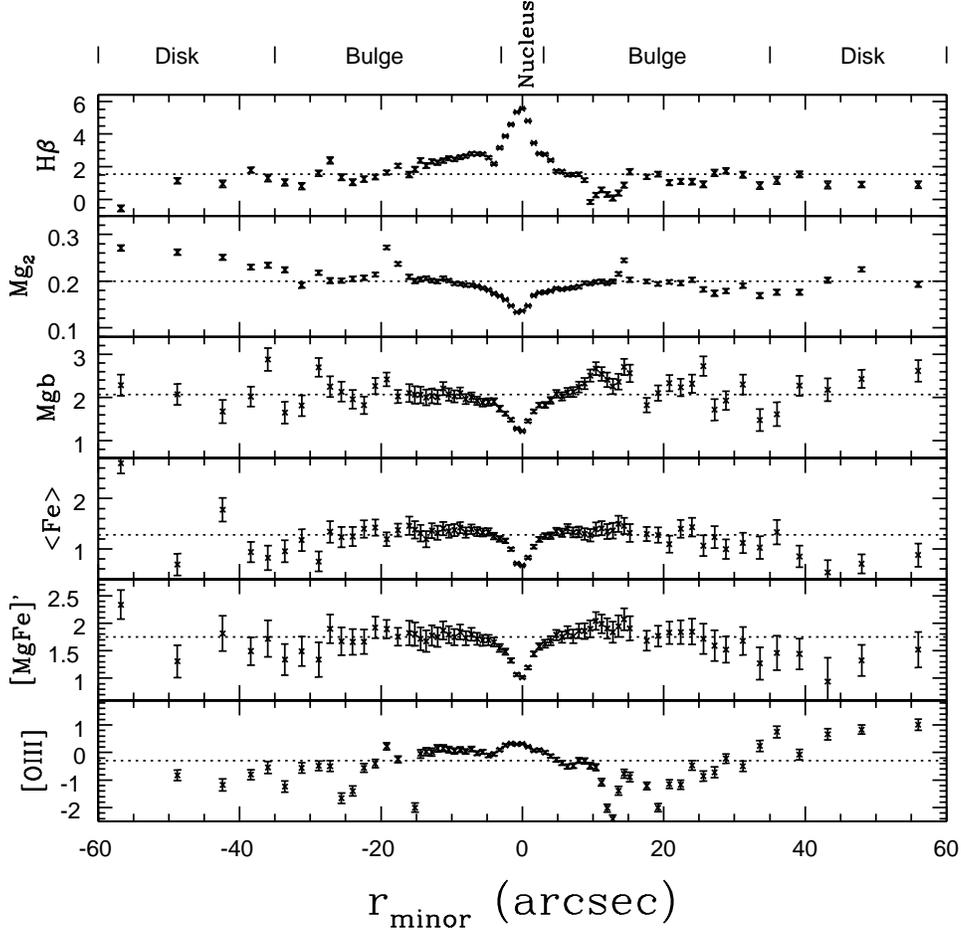}
\caption{Spectral indices. The radial limits of the disk, bulge, and nucleus 
identified from the [3.6] light profile are indicated at the top 
of the figure. Distances are measured along the minor axis 
from the center of the galaxy -- positive values extend to the north, 
negative values to the south. With the exception of Mg$_2$, which is in 
magnitude units, the indices are equivalent widths in \AA . The error bars show the 
random uncertainties estimated from the signal in the various passbands. 
The median value of each index is indicated by a dotted line.
The dip in H$\beta$ between 10 and 15 arcsec is mirrored by an increase in 
[OIII] emission, and it is likely that H$\beta$ absorption at these radii 
is partially filled by emission. The $<Fe>$ and Mgb indices show more-or-less 
symmetric behaviour throughout the bulge and disk, although the indices 
at r$_{minor} = -57$ arcsec differ markedly from those at other radii. 
The asymmetric distribution of the [OIII]5007 index is consistent with the skewed 
distribution of line emission mapped by McMillan et al. (1994). The high 
[OIII]5007 index at $r_{minor} > 30$ arcsec indicates a region of weak 
or non-existant [OIII] emission, which corresponds roughly to the interior of the 
bubble seen in the McMillan et al. (1994) images.}
\end{figure}

\clearpage

\begin{figure}
\figurenum{4}
\epsscale{1.00}
\plotone{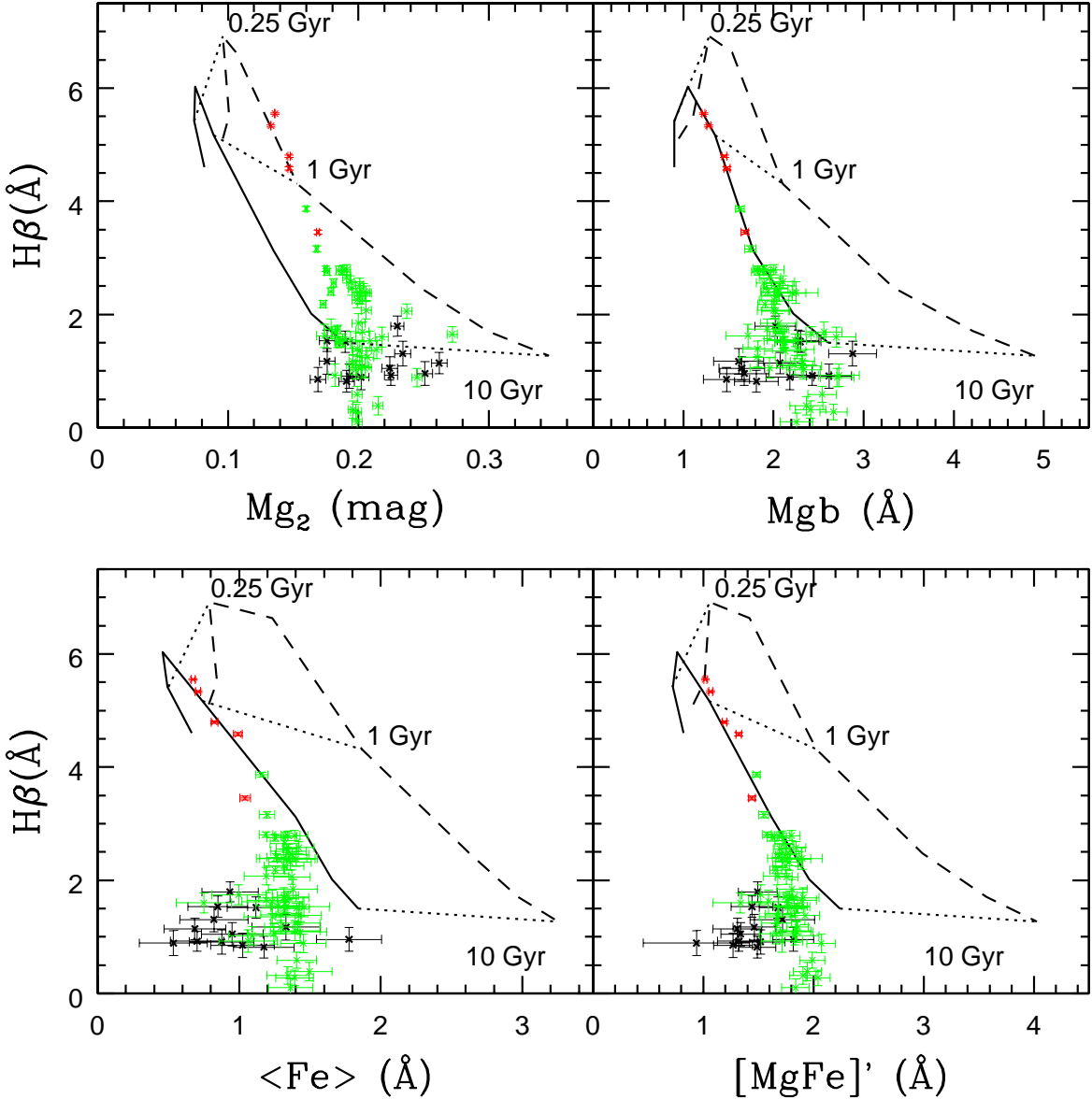}
\caption{Comparisons with model indices. The solid lines show models with 
Z=0.004, while the dashed lines are models with a 
solar metallicity. The dotted lines connect points having the 
same age. A solar chemical mixture is assumed. Indices in the 
nucleus (red), bulge (green), and disk (black) of NGC 5102 are plotted. 
These comparisons suggest that a range of luminosity-weighted ages are sampled, 
from $\sim 0.5$ Gyr in the nucleus to many Gyr in the disk. With the exception of 
the nuclear regions on the H$\beta\ vs$ Mg$_2$ diagram, the models suggest that stars 
in NGC 5102 have sub-solar metallicities, in agreement with the photometric properties 
of resolved RGB stars studied by Davidge (2008). The location of points on the 
H$\beta\ vs$ Mgb and H$\beta\ vs <Fe>$ planes also suggest slightly different 
metallicities for the bulge and disk, as expected if the average [Mg/Fe] 
is non-solar.}
\end{figure}

\clearpage

\begin{figure}
\figurenum{5}
\epsscale{1.00}
\plotone{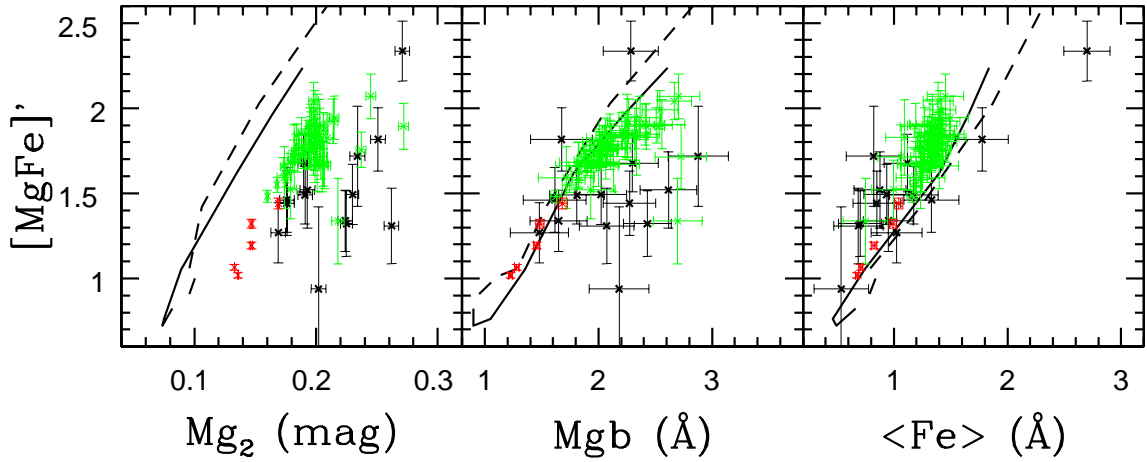}
\caption{Comparisons between models and observations. 
The solid lines are sequences for Z=0.004, while the dashed lines 
assume solar metallicity. Indices are shown for the disk (black), bulge (green), 
and nuclear (red) regions. The NGC 5102 indices fall within the range of values 
predicted by the Z = 0.004 models. That the Mgb and $<Fe>$ points tend to lie on 
different sides of the Z=0.004 model sequence suggests that [Mg/Fe] in NGC 5102 
differs from solar. There is a systematic offset of 
0.03 - 0.04 magnitudes between the models and observations on the [MgFe]' $vs$ Mg$_2$ 
plane. The Mg$_2$ index uses data over a much wider wavelength interval than the 
other indices, and this offset may be tied to the 
relation used to remove the continuum.}
\end{figure}

\clearpage

\begin{figure}
\figurenum{6}
\epsscale{1.00}
\plotone{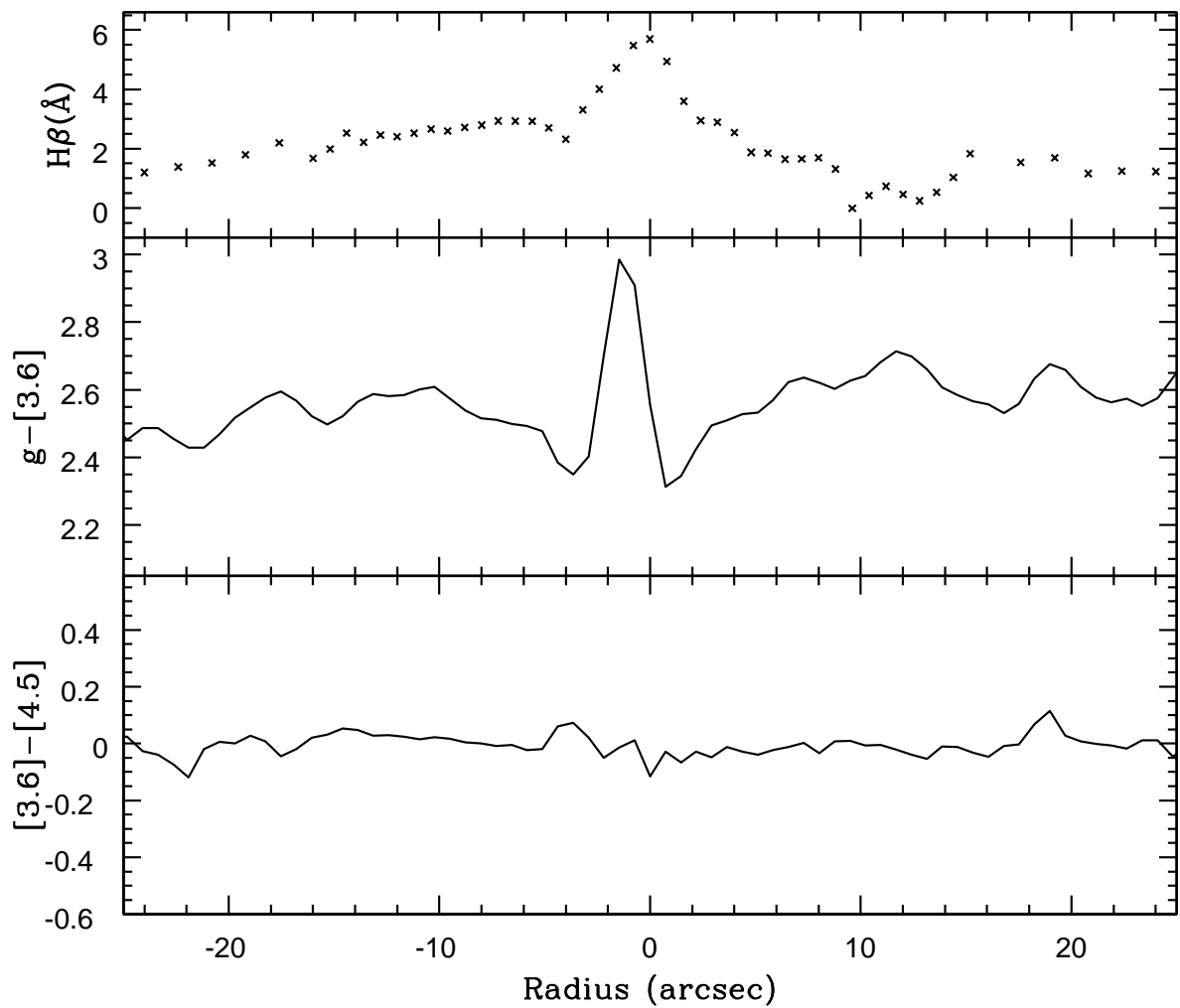}
\caption{Color profiles. The H$\beta$ indices from Figure 3 are re-plotted in 
the top panel. In contrast to the blue color at visible wavelengths, 
the nucleus has a $g'-[3.6]$ color that is redder than the surroundings, and the 
angular extent of the red nucleus matches the region of deep H$\beta$ absorption. 
The red color of the nucleus is due to bright AGB stars that contribute significantly 
to the [3.6] light. While the [3.6]--[4.5] color is 
only weakly sensitive to stellar content, the modest variation 
in this color between 5 and 15 arcsec from the nucleus suggests that the inner 
bulge has an age that is stable to within $\pm 0.4$ dex.}
\end{figure}

\clearpage

\begin{figure}
\figurenum{7}
\epsscale{1.00}
\plotone{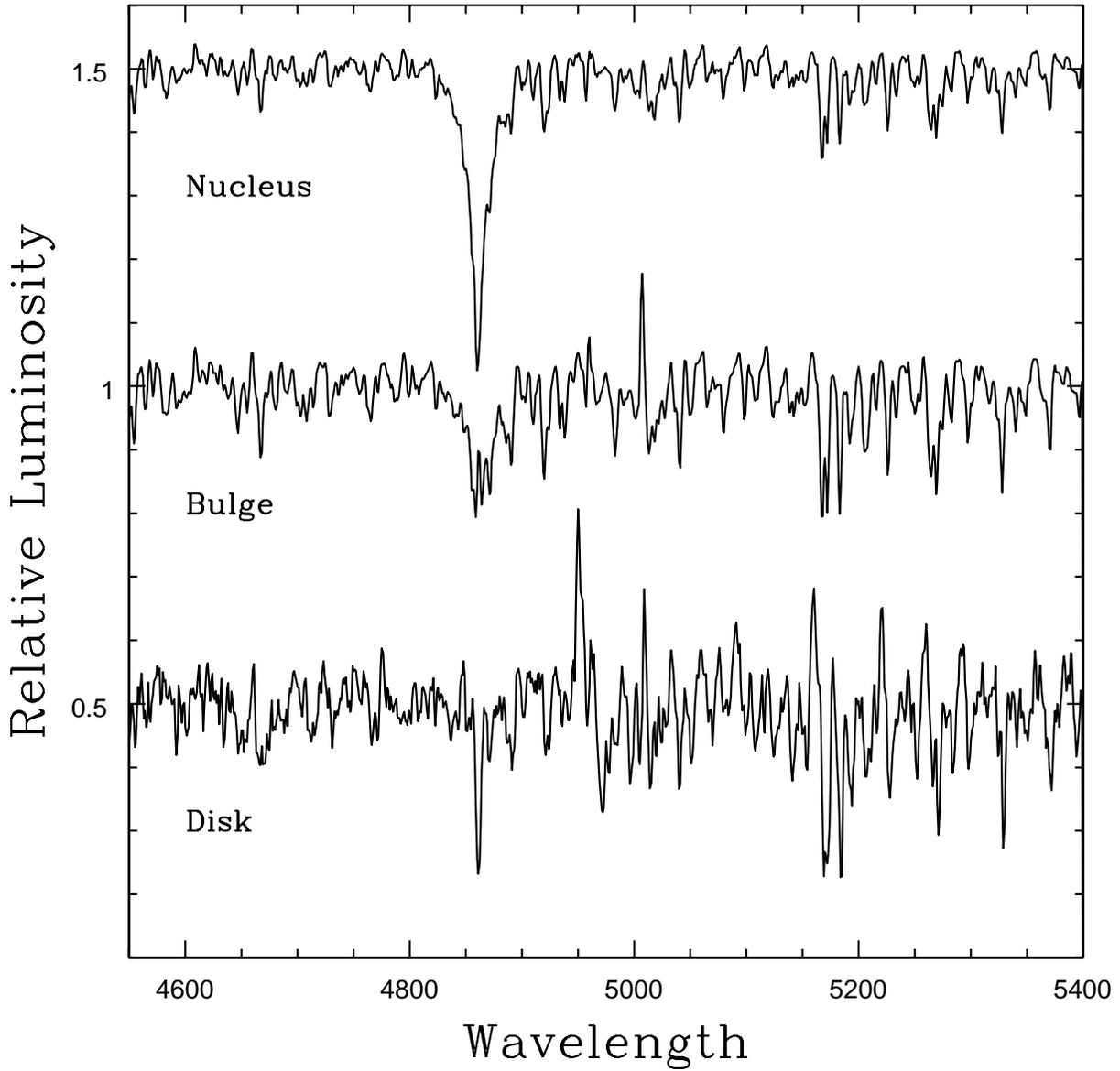}
\caption{Co-added spectra. The spectra were produced by 
summing the extracted spectra in the nucleus, bulge, and disk regions. 
The spectra have been re-sampled to 1 \AA\ to match the models, and 
divided by a high-order continuum function using the 
same fitting parameters that were applied to the model spectra. The results 
have been shifted vertically in this figure for display purposes.}
\end{figure}

\clearpage

\begin{figure}
\figurenum{8}
\epsscale{1.00}
\plotone{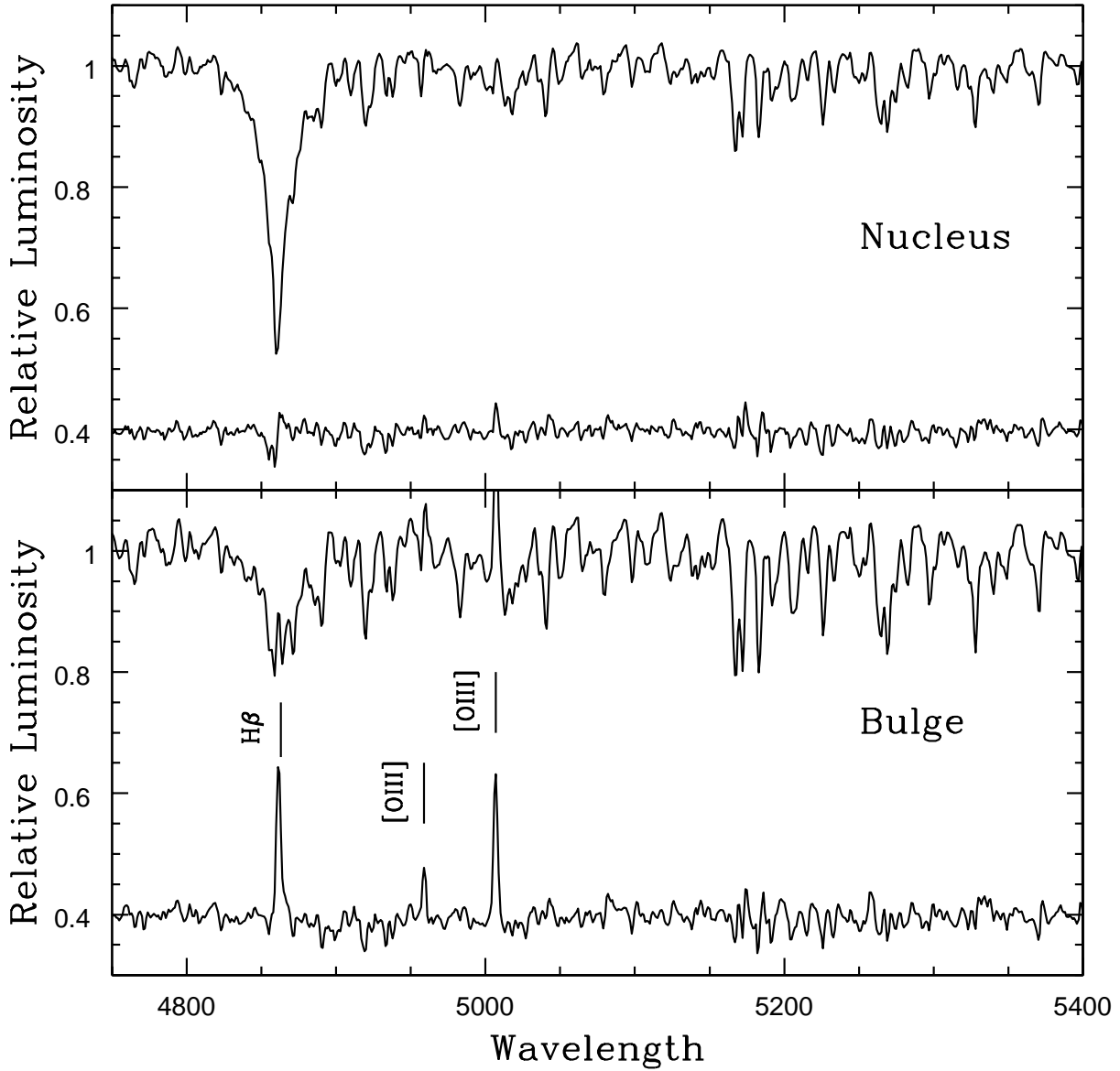}
\caption{Residuals from the SSP models. The spectra of the nucleus and bulge are 
compared with residuals from the 1 Gyr (nucleus) and 2 Gyr (bulge) SSP models. 
The residuals are in the sense observations -- models. Emission lines 
are seen in the residuals, and these are most pronounced in 
the bulge data. The relative intensities of the [OIII] lines 
in the bulge residuals are consistent with theoretical expectations.}
\end{figure}

\end{document}